\documentclass[preprint]{vgtc}                          

\graphicspath{{figures/}{pictures/}{images/}{./}} 

\usepackage{times}                    
\usepackage{enumitem}
\usepackage{wrapfig}

\usepackage{mathptmx}                  

\onlineid{0}

\vgtccategory{Research}

\vgtcinsertpkg

\renewcommand{\sectionautorefname}{\S\kern-1pt}
\renewcommand{\subsectionautorefname}{\S\kern-1pt}
\renewcommand{\subsubsectionautorefname}{\S\kern-1pt}

\title{Multi-User Mobile Augmented Reality for Cardiovascular Surgical Planning}

\newcommand{\tool}[0]{\textsc{ARCollab}}

\newcommand{\authorgap}{\hspace{10pt}}
\renewcommand\footnotemark{}

\author{
    Pratham Mehta\textsuperscript{\textrm 1} %
    \thanks{\textsuperscript{\textrm 1}Georgia Institute of Technology. \{\href{mailto:pratham@gatech.edu}{pratham}$\,\mid\, $
    \href{mailto:rnarayanan39@gatech.edu}{rnarayanan39}$\,\mid\, $
    \newline \textcolor{white}{.} \hspace{11pt}
    \href{mailto:hkaranth3@gatech.edu}{hkaranth3}$\,\mid$
    \href{mailto:alexanderyang@gatech.edu}{alexanderyang}$\,\mid$
    \href{mailto:polo@gatech.edu}{polo}\}@gatech.edu}
    \authorgap
    Rahul Narayanan\textsuperscript{\textrm 1} \authorgap
    Harsha Karanth\textsuperscript{\textrm 1} \authorgap
    Haoyang Yang\textsuperscript{\textrm 1} \authorgap\\
    Timothy C. Slesnick\textsuperscript{\textrm 2} %
    \thanks{\textsuperscript{\textrm 2}Children's Healthcare of Atlanta. \href{mailto:SlesnickT@kidsheart.com}{SlesnickT@kidsheart.com},
    \newline \textcolor{white}{.} \hspace{11pt}
    \href{mailto:Fawwaz.Shaw@choa.org}{Fawwaz.Shaw@choa.org}}  \authorgap
    Fawwaz Shaw\textsuperscript{\textrm 2} \authorgap
    Duen Horng Chau\textsuperscript{\textrm 1} \authorgap
}

\teaser{
  \centering
  \includegraphics[width=\textwidth]{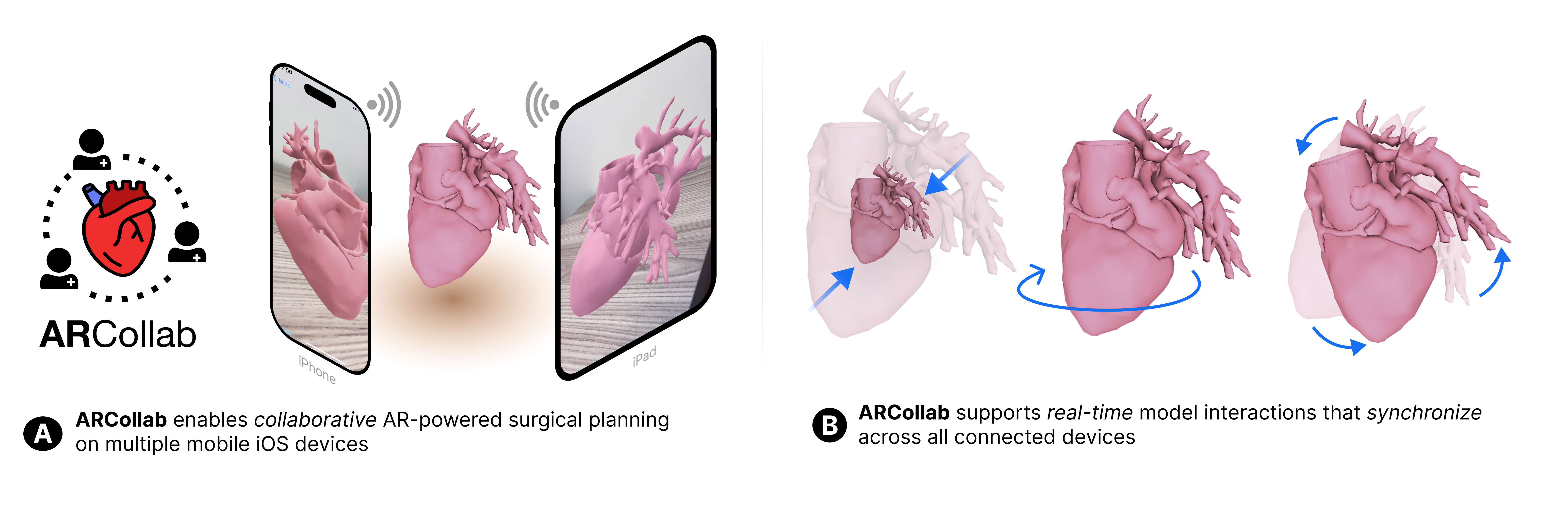}
  \vspace{-3em}
  \caption{(A) \tool{} is a collaborative cardiovascular surgical planning application in mobile augmented reality. Multiple users can join a shared session and view a patient's 3D heart model from different perspectives. (B) \tool{} allows surgeons and cardiologists to collaboratively interact with a 3D heart model in real-time. Changes made on a device update the model's orientation on the other devices in the session.}
  \label{fig:teaser}
}

\abstract{
Collaborative planning for congenital heart diseases typically involves creating physical heart models through 3D printing, which are then examined by both surgeons and cardiologists.
Recent developments in mobile augmented reality (AR) technologies have presented a viable alternative, known for their ease of use and portability.
However, there is still a lack of research examining the utilization of multi-user mobile AR environments to support collaborative planning for cardiovascular surgeries.
We created \tool{}, an iOS AR app designed for enabling multiple surgeons and cardiologists to interact with a patient's 3D heart model in a shared environment. 
\tool{} enables surgeons and cardiologists to import heart models, manipulate them through gestures and collaborate with other users, eliminating the need for fabricating physical heart models. 
Our evaluation of \tool{}'s usability and usefulness  in enhancing collaboration, conducted with three cardiothoracic surgeons and two cardiologists, marks the first human evaluation of a multi-user mobile AR tool for surgical planning.
\tool{} is open-source, available at \url{https://github.com/poloclub/arcollab}.
} 


\begin{document}

\firstsection{Introduction}
\maketitle

The treatment of \textit{congenital heart diseases} (CHDs) necessitates a comprehensive understanding of each patient’s cardiovascular anatomy, achieved through pre-surgical planning sessions. 
During these sessions, surgeons and cardiologists collaborate to create a three-dimensional (3D) printout of the patient's heart. Utilizing \textit{magnetic resonance imaging} (MRI) and \textit{computed tomography} (CT) scans, cardiologists generate a 3D model that accurately details the heart's morphological features \cite{riggs_3d-printed_2018}. 
In many occasions, this model is sliced across a plane to provide a better visualization of the features of the patient's heart 
and it can be beneficial for planning a safer surgery \cite{riggs_3d-printed_2018,Kappanayil2017ThreedimensionalprintedCP,yoo_3d_2021}. 
\setlength{\columnsep}{5mm}%
\setlength{\intextsep}{4mm}

However, these 3D printed models are not easy to procure. Printing such models can take hours and require advanced printing machinery that is not easily accessible to hospitals \cite{yoo_3d_2021}. Additionally, performing slices results in irreversible physical alterations to the physical artifact, making it challenging for the surgeons to effectively explore different cross-sectional slices, impeding the surgical planning process. This prompts the exploration of alternate extended reality (XR) technology such as virtual reality (VR), mixed reality (MR), and augmented reality (AR) to foster more flexible and collaborative planning environments \cite{Sun2019PersonalizedTP, Gr2018AdaptiveA, Kappanayil2017ThreedimensionalprintedCP, yoo_3d_2021, sun_3d_2022, schott_vrar_2021, zhang_directx-based_2022, dass_augmenting_2018, hirzle_critical_2021}. 

The use of mobile AR in medicine shows strong promise due to its portability and familiar interaction environment with simple gestures \cite{Leo21, yang_evaluating_2022, dass_augmenting_2018}. When evaluating such a mobile AR-based surgical planning tool, surgeons and cardiologists remarked on the importance of performing slices across different directions makes it easier to visualize cardiovascular anatomy \cite{yang_evaluating_2022}. They also emphasized that collaboration is an essential part of the surgical planning process and that none of their existing tools supported that \cite{yang_evaluating_2022}. 

To address the above research gaps, our work \textbf{contributes}:

\begin{enumerate}[topsep=2pt, itemsep=0mm, parsep=3pt, leftmargin=10pt]

    \item \textbf{\tool{}, a first multi-user mobile AR-assisted surgical planning tool.} 
    Developed in partnership with cardiologists and cardiothoracic surgeons at \textit{Children's Healthcare of Atlanta} (CHOA), \tool{} enables collaborative cardiovascular surgical planning by allowing multiple doctors to inspect and interact with a 3D heart model in a shared AR space \cite{mehta2024arcollab}. (\autoref{sec:system})
    \tool{} presents a novel design that incorporates Apple's peer-to-peer network (\autoref{subsec:group}) and advanced AR frameworks, facilitating novel slicing operations crucial for detailed surgical planning \cite{yang_evaluating_2022}. 
    To the best of our knowledge, \tool{} is the \textit{first such multi-user iOS application} allowing multiple surgeons and cardiologists to perform omni-directional slicing to the heart (\autoref{subsec:slice}). Fig. \ref{fig:teaser}A illustrates how multiple surgeons and cardiologists would collaborate, each with their own device to collaboratively inspect the heart from multiple perspectives, through real-time transformations such as scaling and rotation (Fig. \ref{fig:teaser}B).
    
    \item \textbf{Reflection and design lessons drawn from human evaluation
with five medical experts.} 
    This work marks the first evaluation of a multi-user mobile AR tool for surgical planning, to the best of our knowledge.
     Three cardiothoracic surgeons and two cardiologists assessed \tool{}'s usability and usefulness,
     demonstrating the ease of use of its  novel features and their potential in improving collaboration. (\autoref{sec:evaluation})
        
    \item \textbf{Streamlining iOS mobile deployment to facilitate broader technology access}. Since \tool{} is built natively on iOS, it can leverage not just the portability and ubiquity of iOS devices, but also their ability to recognize a wide range of gestures and interactions that are intuitive and easy-to-use \cite{dass_augmenting_2018}. 
    By developing within the iOS ecosystem, our design enables the use of Apple's TestFlight service for rapid deployment and iterative testing
    \cite{yang_evaluating_2022}.
    \tool{} is open-source, available at \url{https://github.com/poloclub/arcollab}. (\autoref{sec:system})

\end{enumerate}

\section{Related Works}

\textbf{Collaboration through Physical Models}.
The collaborative surgical planning process begins with the extraction of medical imaging data using \textit{Data Imaging and Communications in Medicine} (DICOM) software \cite{Marro2016ThreeDimensionalPA, Gr2018AdaptiveA}. Cardiologists then create patient-specific heart models using 3D reconstruction and segmentation algorithms \cite{Sundararaghavan2017ThreedimensionalprintedCP,Basso2021ThreeDimensionalPrintedHP}. These models are typically visualized through 3D printing, which allows for detailed examination but is time-consuming and resource-intensive \cite{Batteux20193DPrintedMF,Deakyne2019VirtualPC, Sun2019PersonalizedTP, Gr2018AdaptiveA, Kappanayil2017ThreedimensionalprintedCP, yoo_3d_2021, sun_3d_2022}.

\smallskip \noindent
\textbf{Collaborative XR in Medicine}.
XR tools have recently been proposed to enhance collaborative medical training and planning. Schott et al. developed a multi-user VR and AR interface for liver anatomy education, finding that AR was preferred for its ability to maintain environmental and participant awareness \cite{schott_vrar_2021}. Bork et al. extended this to co-located AR settings, utilizing HoloLens 2 for interactive anatomy learning in medical education, which was well-received by students for its collaborative features \cite{bork_vesarlius_2019}. However, challenges persist with head-mounted displays (HMDs), including discomfort and complex gesture interactions \cite{dass_augmenting_2018, hirzle_critical_2021}. 

\smallskip \noindent
\textbf{Collaborative Mobile AR}.
Mobile AR offers a more intuitive and user-friendly alternative for collaborative medical activities, supporting simple gesture interactions that suit the mobility needs of surgeons and cardiologists \cite{dass_augmenting_2018, Guerrero2006}. Grandi, et al. demonstrated that a handheld 3D user interface could enhance teamwork, improving both accuracy and efficiency in collaborative tasks \cite{Grandi2017DesignAE}. Similarly, Wells, et al. developed a basic mobile AR interface that allows co-located users to view and modify the same virtual objects. The system was effective in supporting collaborative tasks, highlighting the potential of mobile AR in enhancing co-located collaborative workflows\cite{Wells2020CollabAR}, such as surgical planning.

\section{Design Goals}
\label{sec:goals}
We have been collaborating closely with  cardiothoracic surgeons and cardiologists from CHOA over the past two years, iteratively designing and developing \tool{} through regular discussion and consultation; we identified four design goals:

\begin{enumerate}[topsep=1pt, itemsep=2mm, parsep=1pt, leftmargin=19pt, label=\textnormal{\textbf{G\arabic*.}}, ref=G\arabic*]

    \item \label{goal:collab} 
    \textbf{Enable multi-user collaboration in a shared AR space.} Mobile AR significantly enhances co-located collaboration in contexts such as surgical planning \cite{Grandi2017DesignAE,Wells2020CollabAR}. 
    Currently, there is no mobile AR tool for multiple surgeons and cardiologists to collaboratively manipulate a heart model in such an environment. 
    We aim to develop a tool that enables 
    this
    to facilitate collaboration 
    for cardiothoracic surgeons and cardiologists.

    \item \label{goal:portable} 
    \textbf{Support portable usage.} 
    While XR and AR headsets offer immersive 3D model visualization, their utility is hindered by physical form factors, high costs, and poor portability \cite{dass_augmenting_2018,hirzle_critical_2021}. Surgeons and cardiologists are frequently on the move, relying on mobile devices for quick information access. 
    This professional demand for powerful yet portable technology motivates our focus on developing our tool to run on any iOS devices rather than HMDs.

    \item \label{goal:familiar} 
    \textbf{Design easy-to-use user interface with familiar gestures.} 
    The complexity of novel interactions in AR tools often results in a steep learning curve, hindering their  adoption \cite{dass_augmenting_2018}. 
    To address this, our aim is to use widely recognized finger gestures---such as panning, tapping, pinching, and spreading---to facilitate ease of use and promote rapid adoption. This approach will enable surgeons and cardiologists to seamlessly incorporate \tool{} into their practice.

    \item \label{goal:sync} 
    \textbf{Ensure consistent synchronization across multiple devices.}  
    For effective collaboration in surgical planning, it is crucial that all participants view identical updates to the heart models in real-time. 
    Therefore, we design \tool{} to  share only the minimum amount of information across the network, aiming to minimize latency.

\end{enumerate}
\section{System Design and Implementation}
\label{sec:system}

\tool{} is developed with SwiftUI, RealityKit, and ARKit. 

\subsection{Establishing a group session (\ref{goal:collab}, \ref{goal:portable}, \ref{goal:sync})}
\label{subsec:group}
When a user opens the application, the device prompts them to enter their name and proceed to the Devices List, where they can view other nearby available devices to connect with. Multiple users can establish a shared session by tapping on their respective names. This is achieved using Apple’s Bonjour services, a zero-configuration network protocol that connects external devices to a device’s network without the need for assigning specific IP addresses or entering addresses manually. In conjunction with Apple's Multipeer Connectivity framework, the app allows users to detect other devices and connect with them using Wi-Fi if they are on the same network, and Bluetooth if they are not.

\begin{figure}[h]
    \begin{center}
        \includegraphics[width = 0.5\textwidth]{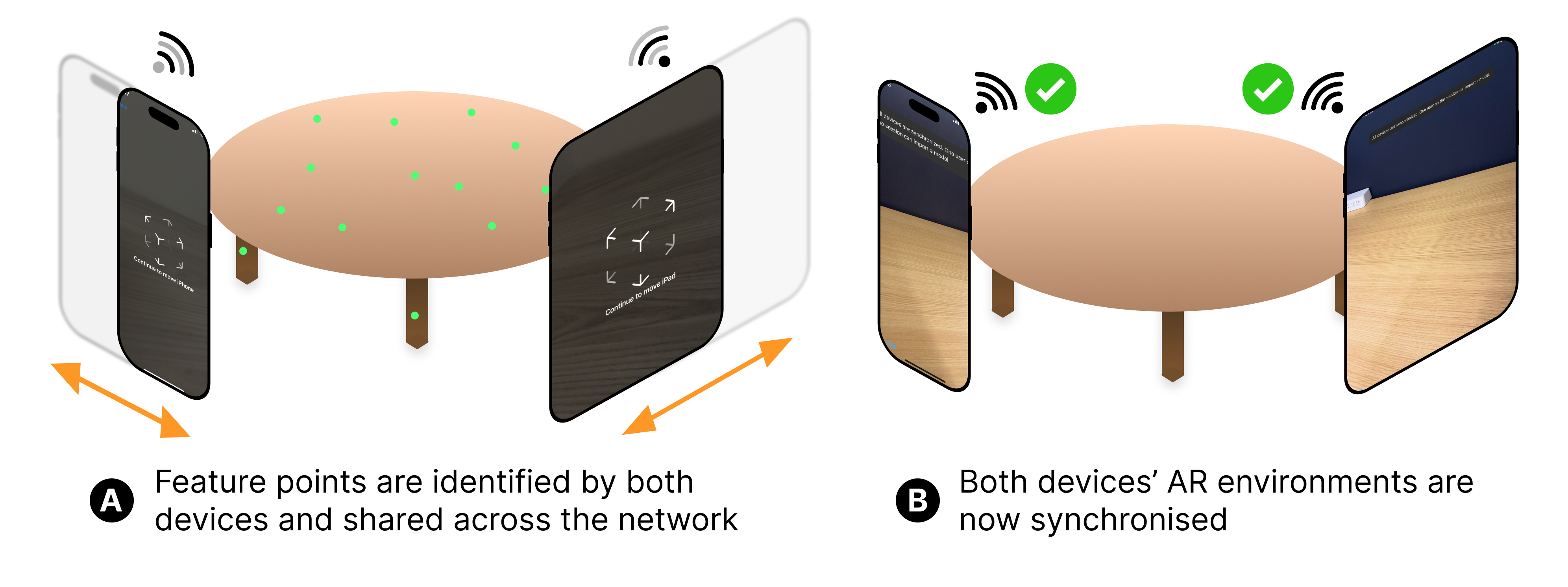}
  \caption{Devices scan environment to detect feature points}
  \label{fig:connectivity}
    \end{center}
\end{figure}

\begin{figure*}
    \centering
    \includegraphics[width=\textwidth]{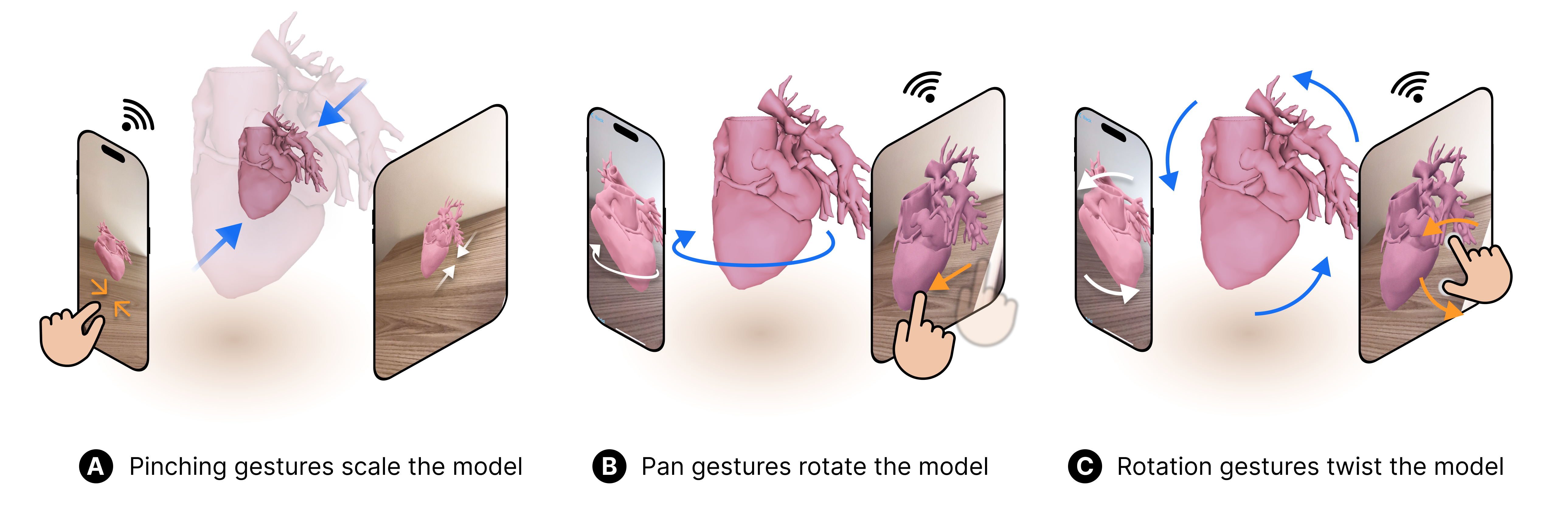}
    \vspace{-5mm}
    \caption{Gestures are used to manipulate the model. The orange arrows represent the gesture performed on one device, the white arrows represent resultant movement of the heart on other devices in the session, and the blue arrows represent actual manipulation of the model. (A) Pinch gesture enables the user to scale the heart model up or down. (B) Panning across the screen triggers rotation of the heart in the direction of the pan. (C) Two-finger rotation on the screen allows the user to rotate the heart around the axis pointing out of the camera perpendicular to the device.}
    \label{fig:interactions}
\end{figure*}

After selecting a set of peers, any user can initiate the session, which navigates all connected devices to the AR View. Users scan the same physical space around them, enabling devices to detect common feature points as indicated in Fig. \ref{fig:connectivity}. Feature points are notable environmental elements that a device can consistently track across frames, such as the corner of a wooden table. These points are significant as they provide reliable markers for the device’s spatial tracking algorithms.

\subsection{Multi-User Model Viewing (\ref{goal:collab}, \ref{goal:familiar})}
\label{subsec:view}

After scanning and calibrating the environment, a surgeon can import a model into \tool{}. The model is projected into the AR environment through raycasting. In this process, the device casts a ray from the center of the screen to detect horizontal planes in front of it. The model is then anchored to the plane, triggering it to render on other devices as well.

Once the model appears on-screen, \tool{} allows  surgeons to interact with it using familiar gestures (\ref{goal:familiar}). 
Scaling the heart is achieved through a pinch gesture using two fingers (\autoref{fig:interactions}A),
rotating is performed through a pan gesture using one finger (\autoref{fig:interactions}B), and twisting the model is done with a rotation gesture using two fingers (\autoref{fig:interactions}C). As a collaborative, multi-user tool, \tool{} supports real-time synchronization of the model across all connected devices (\ref{goal:sync}).
We considered several methods to achieve this synchronization, including sending the entire transformation matrix of the model when any gesture is performed. However, we found this approach to be both resource-heavy and inefficient, resulting in redundant information and causing lag for users. Instead, we opted to  create separate encapsulating data structures for each type of transformation (rotation, scale, twist) and send them individually based on the type of gesture performed. This method proved to be the most efficient during our testing. 

\subsection{Novel multi-user omni-directional slicing (\ref{goal:collab}, \ref{goal:familiar})}
\label{subsec:slice}


Conventionally, surgeons slice a 3D-printed physical heart model along a slicing plane to access the model's internal structures, aiding their understanding of complex surgical morphology and preparation for surgical procedures \cite{Kappanayil2017ThreedimensionalprintedCP}.
According to our collaborating medical experts at CHOA, surgeons can, for instance, position the plane across the ventricular septum or orient it across the left ventricular outflow tract, enabling simultaneous visualization of a ventricular septal defect and the aortic valve, which is beneficial for addressing certain complex defects. However, this uni-directional slicing approach can only generate a limited number of cross-sectional views per physical artifact due to the irreversible nature of the slice \cite{Sun2019PersonalizedTP}. 

\subsubsection{Double-Sided Polygon Rendering (\ref{goal:portable})}

\begin{figure}[h]
    \begin{center}
        \includegraphics[width = 0.3\textwidth]{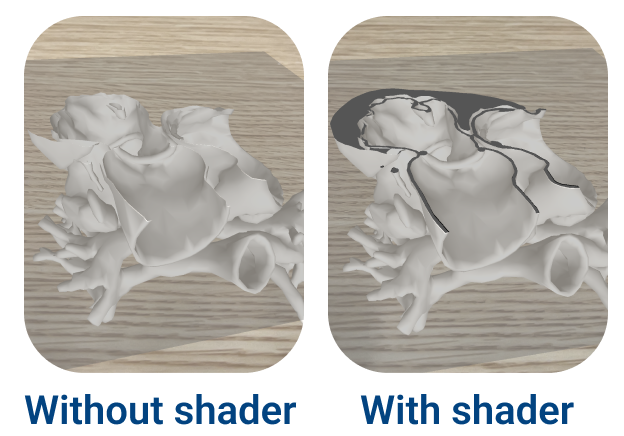}
\vspace{-2mm}
  \caption{Our approach renders the ``inner'' surface of the heart in a darker color to give the appearance of ``filling'' hollow regions.}
\vspace{-5mm}
  \label{fig:shaders}
    \end{center}
\end{figure}

To support multi-user collaborative model slicing in mobile AR, we explored RealityKit, Apple's newest 3D rendering framework. To support custom rendering interactions, RealityKit employs the Metal Shading Language (MSL). By default, MSL only renders one face of the two-dimensional (2D) polygons making up the 3D file, implying that viewing a polygon from the other side will result in empty space as it is not rendered. To replicate reality accurately, both faces of every polygon must be rendered. While RealityKit supports this, it does not allow one side of the polygon to be colored differently from the other, which is crucial for visualizing hollow chambers and tubes in the heart. By shading the ``inner'' surface of the heart with a darker non-reflective color, it appears as if it is ``filling'' the cross-sectional regions intersected by the slicing plane as shown in Fig. \ref{fig:shaders}.

After careful study of MSL and RealityKit, we discovered that we could solve this problem by rendering two separate, identical models, each representing a face of the polygon. Then, we linked them to their own shaders so that the ``inner'' face was colored darker than the other face. With this approach, surgeons and cardiologists would be able to better visualize hollow spaces in the heart model while performing slices. 

\subsubsection{Updating Plane Equation for Model Slicing (\ref{goal:sync})}
Model slicing is not natively supported in RealityKit. To enable this in a multi-user environment, we explored MSL and discovered that we could conditionally render a polygon using surface shaders. We kept track of an equation for the slicing plane of the form \verb|ax+by+cz=d|,  computing the euclidean distance from the plane for every polygon. Then, we rendered every polygon with a non-negative distance value. However, a problem arose with updating this plane equation with finger gestures, as RealityKit stores orientation as a quaternion while Metal only supports the normal form (\verb|ax+by+cz=d|). By exploring the capability of RealityKit, we realized that we could natively convert the quaternion to a 3D transformation matrix. To account for the constant term in the planar equation, we augmented this matrix to four-dimensions by adding a 1 on the diagonal, then multiplied it with the planar equation to obtain the new equation. To ensure that all devices could view the new slice appropriately, the updated plane equation was shared across the network with every gesture, and each device performed the slice on their end.

\subsection{Saving and loading model state (\ref{goal:sync})}
\label{subsec:save}
At any point, the user can click on the \textit{download} icon to save the state of the heart. This action keeps track of the orientation and scale of the heart along with the equation of the plane, triggering other devices to do the same. Then, when the button is used to restore the last saved state, it resets the heart's properties to the saved properties and triggers all devices on the network to do the same.

\section{Usability Evaluation with Medical Experts}
\label{sec:evaluation}
We collaborated with CHOA to recruit five medical experts -- two cardiologists and three surgeons -- who regularly perform cardiovascular surgical procedures 
to evaluate the usability of \tool{}.
CHOA provided the 3D heart model used in the study; the model was constructed by CHOA using a patient's de-identified medical imaging data. 
The study was approved by Georgia Tech’s IRB, with data collection in accordance with institute policies.

\subsection{Procedure}

The user study took place over two days in person, held in a classroom at CHOA. 
Based on the participants' scheduling availability, the first day involved two cardiologists and a cardiothoracic surgeon, while the second day involved two cardiothoracic surgeons. All sessions were audio recorded, and screen recordings were made of all devices used in the study. Each participant signed a consent form before the study commenced.
We provided each participant with an iPad Pro with the latest iOS version and the \tool{} application installed. 
The participants received a brief tutorial, which consisted of diagrams highlighting key features of \tool{}: 
(1) establishing a group session, 
(2) model import, 
(3) model interactions, 
(4) slicing interactions, and 
(5) saving and loading.
Subsequently, participants initiated a group session to use \tool{} for  surgical planning discussions and collaboration.
They were encouraged to verbalize their thoughts as they worked together.
The study concluded with 
a brief questionnaire assessing the usability of the features.

\subsection{Results and Reflection}
The questionnaire comprised 14 questions about the usability and usefulness of \tool{}'s features, with participants asked to rate each question on a 7-point Likert scale (7 being ``strongly agree'' and 1 being ``strongly disagree''). It concluded with open-ended questions soliciting general feedback and identification of top features. We summarized the participants' feedback and our discoveries into the following categories:

\begin{figure}
    \centering
    \includegraphics[width=0.47\textwidth]{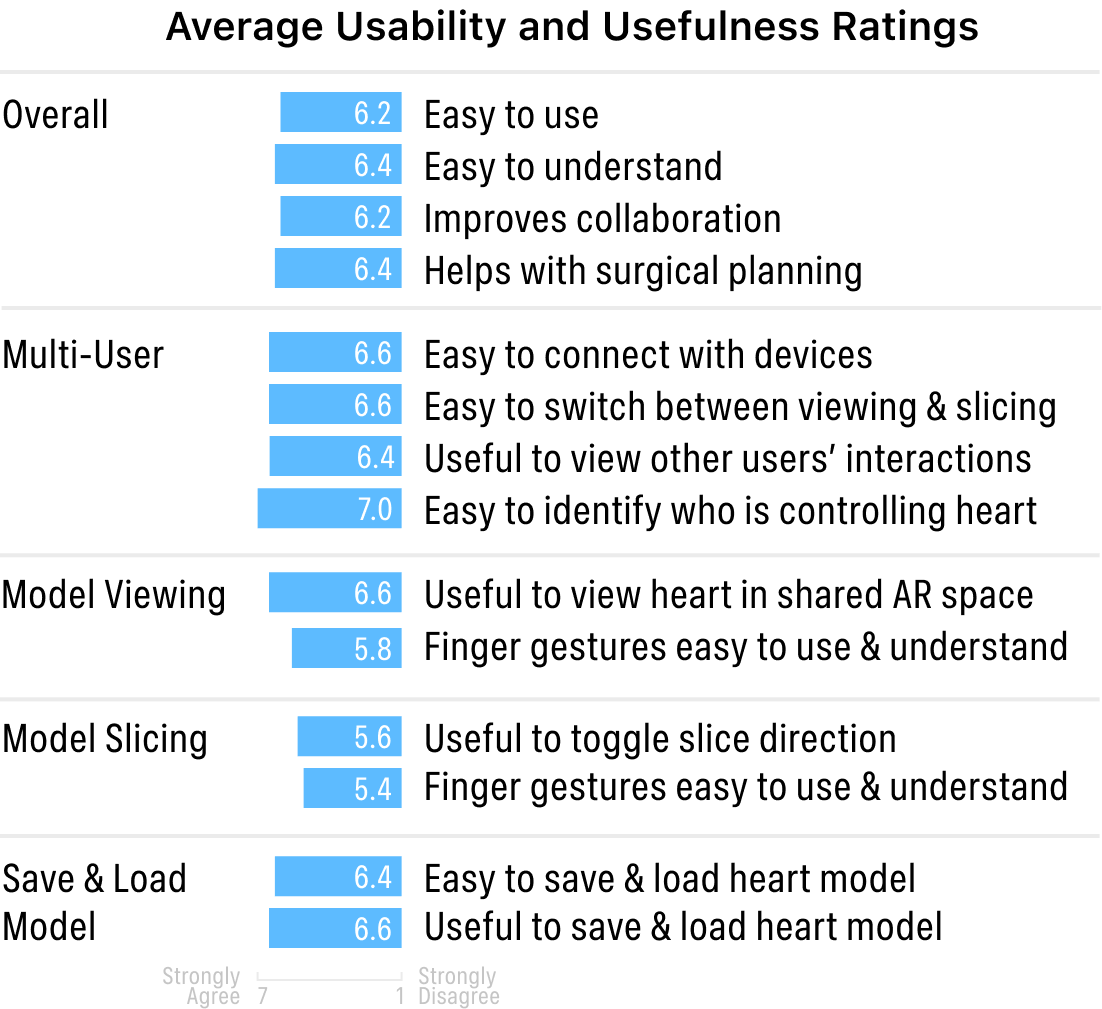}
    \caption{Average ratings about \tool{} from the 2 cardiologists and 3 surgeons who participated in the study. All participants found \tool{} to facilitate collaborative surgical planning.}
    \label{fig:results}
    \vspace{-0.8em}
\end{figure}

\smallskip \noindent
\textbf{New collaboration capabilities enabled by \tool{} are easy to use and facilitate surgical planning.} 
The unique challenge of collaborative surgical planning drove \tool{}'s design decisions, including the adoption of mobile AR to match the mobility needs of surgeons and cardiologists (\ref{goal:portable}), and their familiarity with mobile devices. 
\tool{}'s novel multi-user collaborative features, such as viewing, slicing, and saving,
were all highly-rated for ease of use and understanding (Fig \ref{fig:results}), indicating their potential to facilitate collaboration in surgical planning (\ref{goal:collab}). One participant noted, ``the gestures are easy to pick up. Easier than the Oculus.''  Another participant said, ``I can foresee this being used in planning.''

\smallskip \noindent
\textbf{Anchoring the model to a physical space is important for better interaction.} 
We observed participants comfortably moved around the physical space while interacting with the model. They noted the utility of having the model anchored to a particular spot as it ``is useful to see things [from] different perspectives'' (\ref{goal:sync}). 
One participant also commented, ``[the] ability to walk [around] is intuitive'' while others noted the ``ability to walk around the model'' and ``being able to look at [the] model at same time'' as their top features. 
Participants found it very useful to view the heart in a shared AR space, with an average rating of 6.6 as shown in Fig. \ref{fig:results} (\ref{goal:collab}).

\smallskip \noindent
\textbf{Portability and real-time interaction crucial for collaborative surgical planning.} Our design decision to implement \tool{} in mobile AR was driven by the portability of such tools, reflected in the high average ranking of 6.2 
for ``\tool{} improves collaboration'' (\ref{goal:portable}). 
One participant noted for their top feature, ``mobility–don't require expensive device'' and when giving an example of its convenience, another doctor said, ``If we want to physically interact, we need to 3D print. With a tool like this, we can just import and look through [it] collaboratively.''
A participant also suggested, ``having an iPad or iPhone holder to keep the device stable'' would improve the user experience.
With the portability of mobile devices also comes the familiarity of finger gestures.
We were glad to see that participants felt comfortable with them, with average scores of 5.8 and 5.4 for viewing and slicing gestures, respectively (\ref{goal:familiar}). 

\section{Conclusion and Future Work}

We have presented \tool{}, an iOS AR application that enables multi-user cardiovascular surgical planning in a shared physical space. Through a usability evaluation with medical experts, we demonstrate the efficacy of \tool{} in helping surgeons and cardiologists better collaborate with each other and interact with the model during the planning phase. 

Based on feedback, there are a few ways \tool{} can be improved. For example, by adding support for remote collaboration, \tool{} can better assist doctors in collaborating when not in a shared physical space. Additionally, extending the save feature to support multiple states would further improve collaboration.

\bibliographystyle{abbrv-doi}

\bibliography{main}
\end{document}